# Solving One-Electron Systems in a Novel Gaussian-Sinc Mixed Basis Set

Jonathan L. Jerke, Young Lee and C. J. Tymczak

*Department of Physics, Texas Southern University, Houston, TX 77004, USA*

A novel Gaussian-*Sinc* mixed basis set for the calculation of the one-electron electronic structure within a uniform magnetic field in three dimensions is presented. The one-electron system is used to demonstrate the utility of this new methodology and is a first step in laying the foundation for further development of many-electron atomic and molecular methodology. It is shown in this manuscript how to effectively calculate all basis set integrals, which includes the mixed Gaussian-*Sinc* integrals, with a fast and accurate method. The *Sinc* basis is invariant to the choice of the position of the Coulomb potential, as opposed to traditional grid based methods. This invariance guarantees that the choice of the grids origin has no effect on the electronic structure calculation. This is because the Coulomb potential is treated properly in this methodology, as opposed to DVR methodologies. The off-diagonal terms are sparse but very important around the Coulomb singularity. In general, five to six significant digits of accuracy on all converged results without the linear dependency problems of the Gaussian methodologies are achievable. This methodology is applied to calculate the ground state energy of H atom, $H_2^+$ ion and $H_3^{2+}$ ion in magnetic fields up to a magnetic field strength of $2.35 \times 10^{13}$ G (10,000 au). From these calculations it is shown that $H_3^{2+}$ ion is unstable without relativistic considerations.

## I. INTRODUCTION

A fundamental problem with electronic structure theory is the accuracy of the basis sets that are utilized [1]. Electronic structure basis set methodologies are usually focused on accurately representing either atomic valence states or atomic core states. Core focused methodologies can achieve unprecedented accuracy with the atomic core states with localized basis elements, e.g. Gaussian Type Orbitals (GTO) [2], but fail with valance states. In contrast, valence methodologies that use broad semi- to delocalized basis functions to capture the intricate inter-atom bonding orbitals will achieve high accuracy with bonding orbitals, e.g. Plane Waves [3], but usually fail to capture the highly localized core orbitals. Ideally one would like to be able to move between these two basis themes in a unified construction.

In general, core bases sets require high angular symmetry basis functions in order to span the molecular arrangements of electrons found in molecular and condensed matter systems. However, these high angular symmetry basis functions have two difficulties: a) bonding orbitals can have a symmetry not reflected in the angular symmetries of the core focused bias set used leading to an ill-conditioned expansion; and b) calculation of angular symmetry type basis functions scale as $O(L^4)$ limiting their utilization because of computational complexity. However, it should be noted that these methods have been very successful in solving first row elements, e.g. Crystal03 [4], FreeON [5], or Gaussian [6] for example, but this can be attributed to the low angular symmetry of the constituent atoms.

The electronic valence structure has been adequately modeled with plane-waves basis functions for the primary purpose of spanning the molecular bonding orbitals [3,7-9]. A plane-wave density functional theory (DFT) like ONETEP has found many successes [10]. However, we have chosen to not proceed with a DFT approximation. Furthermore, one can attain the accuracy of plane waves but with A semi-local basis set. A function does exist that is universal enough to reconstruct both plane-waves and be local enough to define properly all interaction tensor terms. The Sinc function is a semi-local function that is both universal and local. Information theory sets a foundation for the *Sinc* function. We will show in this paper that this replacement to plane-wave is local enough to define all interaction tensor elements perfectly, but with the accuracy and utility of the plane wave methods

The valence structure can be spanned with a regular lattice of *Sinc* functions. The *Sinc* is a name given by Shannon [11] and coworkers for the cardinal sine defined in antiquity. Information theory states that band-limited functions can be perfectly represented using a *Sinc* basis. Therefore, compact valence structure can be determined with a finite basis. There is a long history of the *Sinc* function in computational science. ONETEP also uses a periodic version of the *Sinc* function they call *pSincs*. However, we do not use a Fast Fourier Transform (FFT) to define the kinetic energy, instead we use the matrix representation of the *Sinc* derivative to directly represent the kinetic energy of the wave function without loss of generality. The kinetic energy used is found in the literature as the Discrete Variable Representation (DVR) methodology [9,12-14]. However, DVR fails to achieve a solution for the singular Coulomb potential. One property that the DVR method relies upon to obtain accuracy is that all off-diagonal matrix elements approach zero (in relation to the on-diagonal) as the lattice spacing approaches zero. For most cases this is true with the exception of the singular Coulomb potential, which is the source of the previously mentioned problem. Our new method overcomes this problem and in contrast to the DVR methodologies can solve the Coulomb potential with a variational basis set approach in full three-dimensional space. For a review of the *Sinc* representation as used in computational science see Cattoni [15].

For additional accuracy and to correctly represent the atomic core region, this methodology can mix GTO and *Sinc* bases together to achieve a highly accurate representation. This highly flexible mixed basis set allows the choice to either focus on valance, core, or both at the same time in a unified framework. A new methodology of computing the integrals of the *Sinc* and Gaussian functions with the Coulomb potential is developed, which is described in detail in Section III. In this methodology, we are able to compute all integrals within numerical precision necessary to achieve the variational bound, including all the Gaussian-*Sinc* mixed integrals.

In what follows we demonstrate the utility of the *Sinc* function as a variational basis of electronic structure. The *Sinc* function is semi-local enough to define all interaction and matrix elements. Coupled with a Gaussian basis near the cores of atoms, we can achieve a near perfect construction of valence and core orbitals. We generally achieved five to six significant digits of accuracy on all converged results without the linear dependency problems of the Gaussian methodologies [4-6], which will eventually lead to more powerful linear scaling techniques. This methodology was then applied to calculate the ground state energy of H, $H_2^+$ ion and $H_3^{2+}$ ion in magnetic fields completely up to a magnetic field strength of $2.35 \times 10^{13}$ G (10,000 au). From an extrapolation from low field calculations there is little evidence for stability of the $H_3^{2+}$ ion without relativistic considerations, which is in direct opposition to the conclusions of Turbiner [16].

This paper is organized as follows. In Section II a review and present what is known about the *Sinc* function analytic properties. In Section III, a novel method for constructing the Coulomb integrals necessary to compute efficiently all matrix elements is described. In Section IV exhibits basis definitions and consistencies of this scheme. Section V discusses results of one-electron systems including a pure Gaussian calculation, a pure *Sinc* calculation, and a Gaussian-*Sinc* mixed basis. And finally the analysis of the H atom, $H_2^+$ and the $H_3^{2+}$ in magnetic fields up to 10,000 a.u. is presented.

## II. THEORY: THE *SINC* BASIS

For a complete treatise of the subject of the *Sinc* function see Stenger [17]. An outline follows with useful and important properties of the *Sinc* function for general considerations and utility.

### IIA. Discrete Representation

The *Sinc* function that we consider is normalized and defined as:

$$Sinc(\vartheta) = \frac{1}{\sqrt{d}} \frac{Sin(\pi\vartheta)}{\pi\vartheta}, \quad (2.1)$$

where $\vartheta$ is the dimensionless parameter that has a place of an index or sub index written in the form of $x/d$, and $d$ is the distance between lattice sites in one-dimension. The *Sinc* function also is zero at all lattice sites except its own, in other words $\sqrt{d}\,Sinc(n) = \delta_{n0}$. The *Sinc* function is orthonormal on a regular lattice,

$$\delta_{nm} = \int_{-\infty}^{\infty} Sinc(x/d - n) Sinc(x/d - m) dx. \quad (2.2)$$

The origin of the lattice is chosen to be zero here; transformations between different origins will be discussed in Section II.D. These methods offer a discrete representation where the sampling of the band-limited function is the content of the representation,

$$f_n = \int_{-\infty}^{\infty} F(x) Sinc(x/d - n) dx \qquad f(x) = \sum_{n=-\infty}^{\infty} f_n Sinc(x/d - n)$$

$$f_n = \sqrt{d} \int_{-\pi/d}^{\pi/d} \frac{dk}{2\pi} \tilde{F}(k) e^{iknd}. \quad (2.3)$$

The normal convention for the Fourier transform is used. $f_n$ is equal to the value at $nd$ of the convolution of the function, $F(x)$, with a band-pass at a minimum wavelength of $2d$. A band-limited function is perfectly represented on an appropriate lattice, $f(x) = F(x)$, whereby the sampling of the function equals the coefficients up to an overall norm,

$$f_n = \sqrt{d}\,F(nd). \quad (2.4)$$

The *Sinc* function behaves as a discrete delta function on the projection of the function into the band-pass sampled by Shannon's sampling theorem. Engineers call this a low-pass filter of a function [18]. For utility we name these functions Eikons of the functions, because they are similar to the images of the function being represented. Thus, the coefficients on the lattice are the sampling of the low-pass filter of the function, which is seen clearly in Eq. (2.3). The continuous inner product of Eikons is the same as their sampling. The continuous inner product is

$$\langle F|G\rangle = \int_{-\infty}^{\infty} F^*(x)G(x)dx \qquad (2.5)$$

Where Eikons, *F* and *G*, are specified by *f* and *g*,

$$\langle F|G\rangle = \int_{-\infty}^{\infty} dx \sum_n f_n^* Sinc(x/d - n) \sum_m g_m Sinc(x/d - m)$$
$$= \sum_n f_n^* g_n \qquad (2.6)$$

The convolutions of the *Sinc* functions are always translations of band-limited functions; thereby an integer translation of *Sinc* is a Kronecker delta function on the lattice. These expressions are exact for band-limited functions.

## IIB. Scaling Relationship

The *Sinc* function is a universal scaling function. The scaling relationship is due to their band pass nature. A wider band pass admits a narrower band-pass; therefore one can express a *Sinc* at one scale with a lattice of *Sincs* at a finer scale:

$$Sinc(x/d - n) = \sqrt{d'} \sum_m Sinc(md'/d - n) Sinc(x/d' - m) \qquad d' < d \qquad (2.7)$$

The *Sinc* function is completely expressible in terms of other *Sinc* functions at any sub-scale. This property can be very useful in making some calculations tractable. The *Sinc* function is a unique universal function. Only band-limited functions have a scaling relationship that allows them to be expressed in terms of themselves at different origins and scales. All band-limited functions can be expressed as *Sinc* functions. Therefore, the *Sinc* function is the unique universal scaling function of band-limited functions.

## IIC. Incompleteness

The *Sinc* function is a basis that spans band-limited functions. The inner product with a Sinc is necessarily a projection of the state into the band-limited space; necessarily the *Sinc* basis is not complete. Instead, the semi-completeness improves monotonically with the quality of the lattice,

$$Sinc(\tfrac{x-x'}{d}) = \sqrt{d} \sum_n Sinc(x'/d - n) Sinc(x/d - n) \qquad (2.8)$$

This should be contrasted with the standard closure relation for orthogonal basis sets where the left hand side of the equation is a Dirac-delta function. This point has been treated incorrectly in the past. Light [13] claims the *Sinc* basis is complete. This cannot be true because the *Sinc* function only forms a complete basis up to the band edge [17]. This property is the root *approximation* used by the DVR community [9].

**IID. Lattice Derivative:**

The lattice derivative is defined by infinitesimal translations of the Eikon. This procedure leads to semi-local derivative matrix, which when multiplied by the Eikon gives the Eikon of the derivative. One could reconstruct the Eikon continuously into space by using the *Sinc* as the kernel of the interpolation vector. Translations are sampling of the band-limited function at a different origin, and because of Shannon's theorem each sampling is equally good. Referring to Eq. (2.3), the Eikon is translated by, $a$, by the operator,

$$T_{nm}(a) = Sinc(n - m + a/d), \qquad (2.9)$$

where $\mathbf{T}(a)$ operating on the Eikon $\mathbf{f}$ gives,

$$\mathbf{T}(a)f(x) = \sum_m T_{nm}(a) f_m Sinc(x/d - n) = f(x - a) \qquad (2.10)$$

This translation is equivalent to sampling the interpolation of the Eikon at a different origin. We express infinitesimal translations to derive the lattice derivative:

$$\mathbf{D}^{(1)} = \lim_{\varepsilon \to 0} \frac{\mathbf{T}(\varepsilon) - \mathbf{T}(-\varepsilon)}{2\varepsilon} \qquad (2.11)$$

The limit can be taken without respect to any particular Eikon. The derivative matrix $\mathbf{D}$ is then,

$$D^{(1)}_{nm} = \frac{1}{d} \begin{cases} 0, & n = m \\ \frac{(-1)^{n-m}}{n - m}, & n \neq m \end{cases} \qquad (2.12)$$

The expression for $\mathbf{D}$ is semi-local, since an isolated single nonzero lattice element is a *Sinc* function, which is semi-local but zero on the lattice sites. The derivative is necessarily related to the translation operator via:

$$\frac{d\mathbf{T}(x)}{dx} = \mathbf{D}^{(1)} \mathbf{T}(x) \qquad (2.13)$$

This expression shows that the *Sinc* basis can span the derivative of all of its constitute members. We agree with DVR methods [9] that using the exact expression for the second derivative is highly accurate and preferred for the purposes of calculating the derivatives in electronic structure. This expression for the second derivative can be computed via the multiplication of two derivative matrices summed to infinity. Therefore, the second derivative matrix is given,

$$D^{(2)}_{nm} = \frac{1}{d^2} \begin{cases} -\frac{\pi^2}{3}, & n = m \\ -\frac{2(-1)^{n-m}}{(n-m)^2}, & n \neq m \end{cases}. \tag{2.14}$$

Additionally, all these derivative matrices obey the derivative chain rule exactly, as they should

$$\mathbf{D}^{(n+1)} = \mathbf{D}^{(1)}\mathbf{D}^{(n)}. \tag{2.15}$$

**IIE. Commutation Relations**

The commutation relationship of the derivative and the position operator is correct on the lattice up to a factor describing the Fourier component on the edge of Shannon's sampling criterion. This represents the failure of the position operator in a band-limited space more than a true failure of Calculus, since the derivative operator maps the state into the same band-limited space, but the Fourier transform of the position operator is not compact in Fourier space. From Calculus the continuous derivative satisfies the product rule. Specifically on a continuous domain,

$$[\partial/\partial x, x] F(x) = F(x). \tag{2.16}$$

Let us now consider the lattice derivative: The sampling of the derivative is equal to the derivative matrix acting on the sampled function. We propose to check this correspondence by applying the commutator in Eq. (2.16). We derive the commutation relation of the derivative and position operator analytically where $\mathbf{D}^{(1)}$ is the derivative matrix and $\mathbf{X}$ is:

$$X_{nm} = d\, n\, \delta_{nm}. \tag{2.17}$$

From this the commutation relation follows,

$$\begin{aligned}
\left[\mathbf{D}^{(1)}, \mathbf{X}\right]_{ij} &= \sum_k D^{(1)}_{ik} X_{kj} - X_{ik} D^{(1)}_{kj} \\
&= \sum_k \frac{(-1)^{i-k}}{i-k} k \delta_{kj} - i \delta_{ik} \frac{(-1)^{k-j}}{k-j} \\
&= \frac{(-1)^{i-j}}{i-j} j - i \frac{(-1)^{i-j}}{i-j} \\
&= \begin{cases} 0 & i = j \\ -(-1)^{i-j} & i \neq j \end{cases}
\end{aligned} \tag{2.18}$$

Eq. (2.18) is equal to the following form,

$$[\mathbf{D},\mathbf{X}] = \mathbf{I} - \mathbf{A}, \tag{2.19}$$

Where the A matrix is given by,

$$A_{ij} = (-1)^{i-j} \quad (2.20)$$

We refer to **A** as the alternating matrix. The null space of **A** is the number of basis elements minus one. The alternating vector is its span and has a wavelength of Shannon's sampling theorem, specifically *2d*. The alternating matrix stands as a reminder of the sampling theorem, and can only be suppressed with proper encapsulation of the information on the lattice. Alternatively, one can notice that the alternating vector is not translational invariant, so it is an ambiguous object that touches on all possible representations of the same alternating state under translation of the basis.

### III. COULOMB INTERACTION AND MATRIX INTEGRALS

An integral equation is presented to calculate Coulomb terms between Gaussian and Gaussians, *Sincs* and *Sincs*, and Gaussians and *Sincs*. The equations are modular and built on units of the basis in one dimension since these functions are separable in three dimensions. The final expression of the matrix and interaction elements is tractable with available computer technologies. The interaction term is defined by the following wave functions, $H(\vec{r})$ and $G(\vec{r})$,

$$\int d\vec{r}^3 \int d\vec{r}'^3 \frac{G(\vec{r})H(\vec{r}')}{|\vec{r}-\vec{r}'|} \quad (3.1)$$

The Coulomb potential can be expressed in terms of a Gaussian Fourier integral,

$$\begin{aligned} &= \int d^3\vec{r} \int d^3\vec{r}' G(\vec{r})H(\vec{r}') \int \frac{d^3\vec{k}}{(2\pi)^3} \int_0^\infty d\alpha \frac{e^{-\vec{k}^2/(4\alpha^2)-i\vec{k}\cdot(\vec{r}-\vec{r}')}}{\alpha^3} \\ &= 2\pi \int_0^\infty d\alpha \left[ \int_{-\infty}^\infty \frac{d^3\vec{k}}{(2\pi)^3} \frac{e^{-\vec{k}^2/(4\alpha^2)}}{\alpha^3} \tilde{G}(\vec{k})\tilde{H}(-\vec{k}) \right] \end{aligned} \quad (3.2)$$

The Fourier transform of a function is defined in the normal convention,

$$\begin{aligned} \tilde{F}(\vec{k}) &= \int d^3\vec{r} F(\vec{r}) e^{-i\vec{k}\cdot\vec{r}} \\ F(\vec{r}) &= \int \frac{d^3\vec{k}}{(2\pi)^3} \tilde{F}(\vec{k}) e^{i\vec{k}\cdot\vec{r}} \end{aligned} \quad (3.3)$$

Beginning with a six-dimensional integral, one can immediately reduced this to a four-dimensional integral. At all points this expression of the Coulomb integral is exact. We restrict our final calculation to only separable basis elements in three dimensions,

$$F(\vec{r}) = F_x(x)F_y(y)F_z(z) \quad (3.4)$$

Eq. (3.2) separates inside the alpha integral into three one-dimensional integrals,

$$I_i(\alpha) = \left[ \int_{-\infty}^{\infty} dk/(2\pi) \frac{e^{-k^2/(4\alpha^2)}}{\alpha} \tilde{G}_i(k) \tilde{H}_i(-k) \right]$$ (3.5)

The next step is a reduction of this to a one-dimensional "alpha integral" that is the product of one-dimensional Fourier integrals,

$$\int dr^3 \int dr'^3 \frac{G(r)H(r')}{|r-r'|} = 2\pi \int_0^{\infty} d\alpha \, I_x(\alpha) I_y(\alpha) I_z(\alpha)$$ (3.6)

Obviously the original interaction element expression is effectively intractable to numerical integration. However, Eq. (3.6) for separable basis elements is numerically tractable and very efficient. It takes about a second on a single CPU for any term to machine precision. Further improvements in speed can be afforded by analytically expressing Eq. (3.5). The Gaussian-*Sinc* terms are separated inside the "alpha integral", leaving one-dimensional integrals to be computed that can be of the *Sinc-Sinc*, Gaussian-*Sinc*, or Gaussian-Gaussian variety. The *Sinc-Sinc* term has an alternative but equal useful expression,

$$Sinc(x/d - n) Sinc(x/d - m)$$
$$= d \int_{-\pi/d}^{\pi/d} \frac{dk_1}{2\pi} \int_{-\pi/d}^{\pi/d} \frac{dk_2}{2\pi} e^{-i(k_1(x-dn) + k_2(x-dm))}$$ (3.7)

The Gaussian-*Sinc* term is,

$$\left(\tfrac{2\beta}{\pi}\right)^{\tfrac{1}{4}} e^{-\beta(x-x_0)^2} Sinc(x/d - n)$$
$$= \left(\tfrac{2\beta}{\pi}\right)^{\tfrac{1}{4}} e^{-\beta(x-x_0)^2} \sqrt{d} \int_{-\pi/d}^{\pi/d} \frac{dk}{2\pi} e^{-ik(x-dn)}$$ (3.8)

The Gaussian-Gaussian term is also Gaussian,

$$\left(\tfrac{2\beta_1}{\pi}\right)^{\tfrac{1}{4}} \left(\tfrac{2\beta_2}{\pi}\right)^{\tfrac{1}{4}} e^{-\beta_1(x-x_1)^2} e^{-\beta_2(x-x_2)^2}$$ (3.9)

This completes the list of all necessary basis constructions needed to construct all interaction terms in the Gaussian–*Sinc* basis. All corresponding Fourier transforms of the functions listed above can be analytically expressed.

The Fourier transform of the *Sinc-Sinc* term is analytically expressible. A single *Sinc* function has a Fourier Transform of a top hat and a double Sinc function has a Fourier transform of a triangle:

$$\begin{cases} -ie^{iknd}\dfrac{\left(-e^{i(m-n)(kd-\pi)}+e^{i(m-n)\pi}\right)}{2\pi(m-n)}, & 0<k<2\pi/d, \quad m\neq n \\[2mm] ie^{iknd}\dfrac{\left(-e^{i(m-n)(kd+\pi)}+e^{-i(m-n)\pi}\right)}{2\pi(m-n)}, & -2\pi/d<k<0, \quad m\neq n \\[2mm] \dfrac{e^{iknd}\left(2\pi-|kd|\right)}{2\pi}, & -2\pi/d<k<2\pi/d, \quad m=n \end{cases} \quad (3.10)$$

The expression for the Gaussian-*Sinc* Fourier transform is more difficult and presents a larger problem generally. This expression is written in terms of Error functions,

$$2\sqrt{d}\left(\tfrac{\beta}{2\pi}\right)^{1/4} e^{iknd} e^{-\beta(nd-x_0)^2} \left( Erf\left[\tfrac{-k+\pi/d-2i\beta(nd-x_0)}{2\sqrt{\beta}}\right] - Erf\left[\tfrac{-k-\pi/d-2i\beta(nd-x_0)}{2\sqrt{\beta}}\right] \right). \quad (3.11)$$

The ultimate problem with Error functions is that they are not analytically integrable. This presents a problem for further analytic reduction. Analytic expressions are desirable because of the speed and accuracy for which they can be calculated. The analytic expression for the Fourier Transform of a Gaussian-Gaussian term is well known to be a Gaussian. The position of the Gaussian is a beta weighted average and the distance between them has Gaussian suppression,

$$\sqrt{2}\left(\tfrac{\beta_1\beta_2}{(\beta_1+\beta_2)^2}\right)^{1/4} e^{-\tfrac{k^2}{4(\beta_1+\beta_2)}-\tfrac{\beta_1\beta_2(x_1-x_2)^2}{\beta_1+\beta_2}-ik\tfrac{\beta_1 x_1+\beta_2 x_2}{\beta_1+\beta_2}}. \quad (3.12)$$

This is a trivial point in this context, but may be of more general interest to the Gaussian community. This formula allows us to integrate any Gaussian integral. In fact it is straightforward to use Eq. (3.5) to solve for any angular symmetry type of a Gaussian-Gaussian interaction term generally under any condition. One could also imagine different betas in every direction. At this point we can see why plane-wave basis elements are intractable. The Fourier transform of a plane wave is a Dirac delta, and equation Eq. (3.5) prescribes two Dirac Deltas per integral per dimension. Therefore the result is undefined.

An adaptive integral procedure [19] can integrate these integrals at a rate of one element per second. However, this can be substantially speed up for all one-body matrix elements allowing for the very rapid calculation times. We have calculated all matrix elements in this paper with numerical precision of 9 significant digits at a rate of about 1000 per second.

## IV. THE SCHRÖDINGER WAVE EQUATION

### IVA. Gaussian-*Sinc* Basis

The Gaussian basis extends and compliments the *Sinc* basis allowing for a more accurate treatment of the "cusp" of the wave function without the linear dependences of broad Gaussians. The Gaussian basis elements are placed at the atomic centers to accurately span the atomic core structure. The Gaussian exponents are restricted such that inter-atomic overlap is below a defined threshold, significantly simplifying computation because these Gaussian-Gaussian terms from one atom to another can usually be neglected.

**IVB. Upper Bounded**

Without magnetic fields, we are using the true matrix elements of the basis functions mentioned and therefore remain variationally bound. One can prune the Sinc basis without loss of this principle. We do not need to assume periodicity or other boundary conditions. The construction guarantees that the wave function goes to zero at infinity. The *Sinc* function is the proper way to sample the wave function in a digital representation. The *Sinc* function is the interpolation kernel for a sampling of a function. The tricky point is the lattice scale, however it is conceivable to integrate even multi-resolution *Sinc* functions. Wavelets and multi-resolution run parallel in the *Sinc* formalism [ref].

**IVC. Definitions**

**Momentum Operator**: The momentum operator **P** is proportional to the derivative operator; therefore we can define it as,

$$\mathbf{P} = -i\hbar \mathbf{D}^{(1)}. \tag{4.1}$$

**Canonical Commutation Relations**: Using Eq. (2.19), we evaluate the canonical commutations relationships used in Quantum Mechanics as,

$$[\mathbf{X}, \mathbf{P}] = i\hbar(\mathbf{I} - \mathbf{A}). \tag{4.2}$$

**Schrödinger Wave Equation in a Magnetic Field:** From Eq. (2.14) and Eq. (3.1) we construct the non-relativistic Schrodinger equation in a zero magnetic field for quantum mechanical problems in our Eikon method as,

$$\mathbf{H} = \left( \frac{\mathbf{P}^2}{2m} + \mathbf{V} \right). \tag{4.3}$$

To include magnetic fields, we begin with the standard construction,

$$\mathbf{P} \rightarrow \mathbf{P} + \tfrac{e}{c}\mathbf{A}, \tag{4.4}$$

Which give us the Hamiltonian

$$\mathbf{H} = \frac{1}{2m}\left(\mathbf{P} + \tfrac{e}{c}\mathbf{A}\right)^2 + \mathbf{V}. \tag{4.5}$$

For a uniform magnetic field, where $\mathbf{A} = \tfrac{1}{2}\mathbf{r} \times \mathbf{B}$, we can rewrite as for a magnetic field in the z-direction,

$$\mathbf{H} = -\frac{\hbar^2}{2m}\mathbf{D}^{(2)} - \frac{ie\hbar}{2mc}(\mathbf{x}\mathbf{D}_y^{(1)} - \mathbf{y}\mathbf{D}_x^{(1)})B + \frac{e^2}{8mc^2}\left(\mathbf{x}^2 + \mathbf{y}^2\right)B^2 + \mathbf{V}. \tag{4.6}$$

The mixed basis requires an overlap matrix to solve the generalized Eigen-value problem, which has been solved for real and complex Hamiltonians using the LAPACK computational eigen-solver package [20].

## V. ONE-ELECTRON SYSTEMS

### VA. Hydrogen Benchmark Calculation with Pure Gaussians

We have applied our new integral technology to compute the non-interacting eigen-states of Hydrogen using a pure Gaussian basis in order to benchmark the methodology. Previous Gaussian algorithms require recursive and/or many function evaluations to compute matrix element and scale poorly with angular symmetry, $O(L^4)$ computational complexity per matrix element (where there are $O(L^2)$ matrix elements), with angular symmetry types [21]. Our integral methodology allow us to solve for any matrix element directly and computationally very efficiently, with $O(L)$ computational complexity per matrix element. Calculating the non-interacting Eigen-state of Hydrogen demonstrates the quality of a solution of core states in this computational scheme. 100 GTOs are used with estimated error on each matrix element below $10^{-11}$. We used the simple formula for the Gaussian exponents,

$$\beta_i = \alpha_0 \left( \beta_0^{i-1} \right) \tag{5.1}$$

Where $\alpha_0 = 0.025$ and $\beta_0 = 2.0$ with 25 s-type Gaussians and 3x25 p-type Gaussians, a total of 100 GTO primitives. The 1s Eigen-state has an energy of -0.49999999948, the 2s Eigen-state energy was -0.249999999967 and the 2p Eigen-state energy was -0.249999999942. The total calculation took approximately ten seconds of CPU time.

### VB. Calculation of Hydrogen Ground State with *Sincs*

Next, the new *Sinc* methodology is applied to the ground state eigen-energies of Hydrogen. We calculated the Hydrogen system with a radius of 7.5 au from the nucleus in spherically pruned basis of lattice elements with lattice spacing of d=0.5, 1.0, and 2.0 au. The results of this calculation are presented in Table I. Resultant errors found are 0.6% error at d=0.5, 4% error at d=1.0, and 15% error at d=2.0. For reference, we compare this to previous wavelet calculations [22-29]. Wavelets could be considered the state-of-art when it comes to semi-analytical grid basis. As a specific example, consider the results of Tymczak et. al. [30]. Tymczak reports 2% errors on the computation of the same system. We see that Tymczak's sparsely filled $128^3$=2,097,152 grid results were comparable to the one grid, yet the one grid is significantly smaller. Additionally, it was stated in this work that there was a sensitivity of the calculated eigen-energies to the position of the nuclear center in respect to the grid, which would cause significant issues with the calculation of atomic forces using these methods. This problem is rectified with the *Sinc* basis. In Figure 1 we plot a grid of d=0.5 calculation of the probability density centered and translated. There is convergence of the energies between the centered and translation of the one-grid calculations of increasing range of the wave-function. From this we conclude that any variations of the energies are strictly due to edge effects and not any sensitivity of the solution to the position of the singularity. We find it difficult to compare our results to most other calculation because our results do *not* assume any symmetry. Almost every paper solving hydrogen has assumed separation of variables in spherical or cylindrical coordinates, which is irrelevant to almost all molecular systems. For instance, finite element methods [31] and DVR [32] solved Hydrogen to very high accuracy in that symmetry. Such papers are useful as benchmarks for this work, but cannot be compared equally to a calculation conducted in three dimensions. Three-dimensional calculations are much more flexible to molecular configurations and are therefore a much more useful tool in solving the electronic structure without bias.

Additionally, in Figure 2 we plot the error in ground state energy of hydrogen as the grid spacing decreases, as can be seen we obtained exponential convergence with decreasing grid spacing. For comparison, we also show in this figure the convergence of pure Hydrogen in a magnetic field and the $H_2^+$ ion.

**VC. Hydrogen in the Gaussian-*Sinc* Basis**

In this section we extend our calculation of the ground state of Hydrogen using a mixed Gaussian-*Sinc* basis set. To minimize the linear dependence but still achieve a span of the cusp with the Gaussian basis we have restricted the minimum exponent. Using Eq. (5.1), we varied the alpha and beta values with a set number of Gaussians until the energy was minimized. Our results are reported in Table III for the Hydrogen atom using the *Sinc*-Gaussian basis set. As is expected, as we decrease the grid spacing the minimum exponents increases. One problem that this addresses is linear dependence of the basis set. The major contribution to this linear dependence is the Gaussian *Sinc* overlap. This dependence becomes significant when the Gaussian exponent and the grid spacing become comparable. However, with careful optimization we can easilly avoid this issue, as is shown in the table.

**VD. Results for Hydrogen in a Magnetic Field**

In Figures 3 and 4 we plot the cross sections of the ground state wave functions at various magnetic fields strengths. The convergence of the solutions with increasing sampling rate and range of the neighborhood around the atoms are given in Table IV and Table V for 1 au and 10 au magnetic fields respectively. Figure 3 shows the cross sections along all three axes of the wave function at magnetic field of 1 au. The energy of this calculation has been known to be -0.3312 Hartrees [33] and we found -0.33114 Hartrees. At 10 au of magnetic field we plot the best solution in Figure 4. The known calculated energy from reference [33] is 3.2522 Hartrees and our result was 3.25222 Hartrees. Energies remained within tolerance under translation by 0.5 au transverse to the magnetic field.

**VE. Results for the $H_2^+$ ion in the Gaussian-Sinc basis**

We calculated the ground state configuration $H_2^+$ in a Gaussian-*Sinc* basis, as shown in Table VI. From the table we see that we agree with the known energies to an accuracy of six significant digits, where the known energy of the system is reported by Scott et. al. [34] to be -0.60263 Hartree at 1.998 au separation. In this reference the author details the technique of searching a hyper-dimensional trial wave function parameter space to find the minimum energy. The advantage of our method is that we do not need to conduct such a complex multidimensional optimization.

**VF. Result of the $H_2^+$ ion in a Magnetic Field**

We report on the calculation of the $H_2^+$ ion system in a magnetic field of 1 and 10 au. Results shown in Table VII and Table VIII are directly compared to established work on one electrons systems by Turbiner et. al. [16]. Specifically, Our calculation at $45^0$ is of lower energy then previous calculation and is consistent under translation of 0.5 au, which is a validation of the utility of our method.

**VG. Results for the Trigonal $H_3^{2+}$ ion in a Magnetic Field**

In Figure 5, the ground state wave function of trigonal $H_3^{2+}$ ion for a zero magnetic field and a magnetic field of 1 au. Freezing the positions of the atoms, the electron completely leaves the orbit of the third proton with this magnetic field. This is indicative of the instability of the trigonal $H_3^{2+}$ ion in a magnetic field. We also find that the linear $H_3^{2+}$ is always significantly more stable then the trigonal $H_3^{2+}$, therefore we will only consider the linear configuration in the next section for which we will explore these systems in more detail.

## VH. Energetic's in Very High Magnetic Fields

We have conducted a large-scale scan from zero to 10,000 au of magnetic field for the H, $H_2^+$ and $H_3^{2+}$ systems. From this we can conclude $H_3^{2+}$ system is unstable to decay into $H_2^+$ without relativistic considerations. The results are reported in Table IX. In Figure 6 the calculated energies of the one electron species with respect to the energy of Hydrogen within the same field are plotted. Fitting the data to a functional form,

$$F(B) = \frac{a + bB + cB^2}{d + eB},\qquad(5.2)$$

which is also shown in Figure 6. We find that the energies converge to parallel lines with energy difference of approximately 0.2 Hartree for all field predictions up to the relativistic limit. All our calculations agree with reference [16] result to a hundredth of a Hartree for all magnetic fields considered. We disagree with the conclusion of reference [16] that the $H_3^{2+}$ ion becomes the more stable species at high magnetic fields and believe that the incorrect conclusion in this reference is do to the scarcity of data points in the calculation. We warn that calculations above 5,000 au are questionable at best and its better to fit trends below that field value. The difficulty of trans-10,000 au field calculations places very little weight on their final numbers of a bound state to $H_3^{++}$. Whereas a trend in a more trustworthy domain of field with a reliable extrapolation scheme has more weight. No aspect of our analysis biased the result to lead to parallel lines above 10,000 au of field. Even so, there never is a transition to an $H_3^{++}$ ground state. We therefore conclude that $H_3^{++}$ is unstable to decay in this non-relativistic theory.

## VI. CONCLUSION

We have demonstrated the use of a novel Gaussian-Sinc basis set for the calculation of the electronic structure within a uniform magnetic field in three dimensions. With this method, one can achieve four to six significant digits in the ground state energy for any proton configuration. One can extend this technology to nonzero magnetic fields to show that $H_3^{2+}$ is unstable in all magnetic fields without relativistic corrections. Additionally, we are in the process of coding this method to include a many-body description (Hartree-Fock level) that will allow us to study atoms and molecules in ultra-high magnetic fields.

## VII. ACKNOWLEDGEMENTS

The authors would like to acknowledge the support given by Welch Foundation (Grant J-1675), the ARO (Grant W911Nf-13-1-0162), the Texas Southern University High Performance Computing Center (http:/hpcc.tsu.edu/; Grant PHY-1126251) and NSF-CREST CRCN project (Grant HRD-1137732). Dr. Jerke would also like to thank University of Houston for hosting him as a Visiting Assistant Professor while beginning to consider information theory to be the foundation of Quantum Mechanics. We would also like to thank G. Reiter for many fruitful discussions.

## VIII. FIGURES

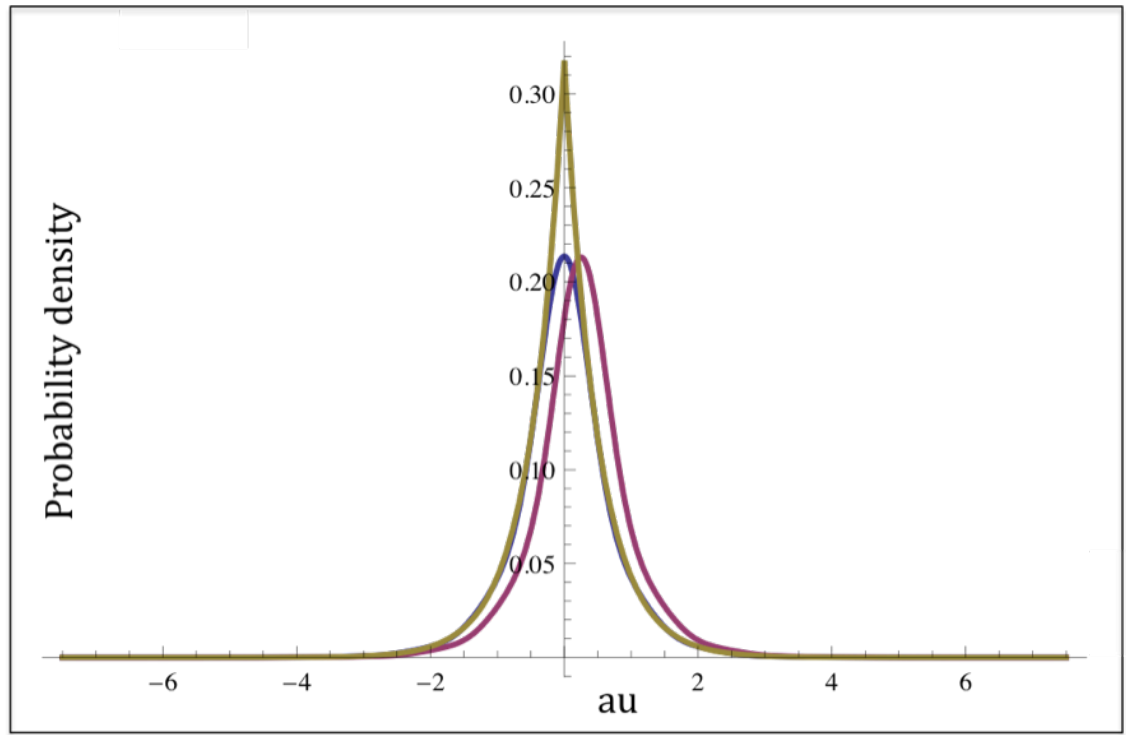

Fig. 1. A pure *Sinc* half-grid calculation of Hydrogen with the analytic solution superimposed in yellow. Translation of the potential successfully leads to translation of the solution's wave function.

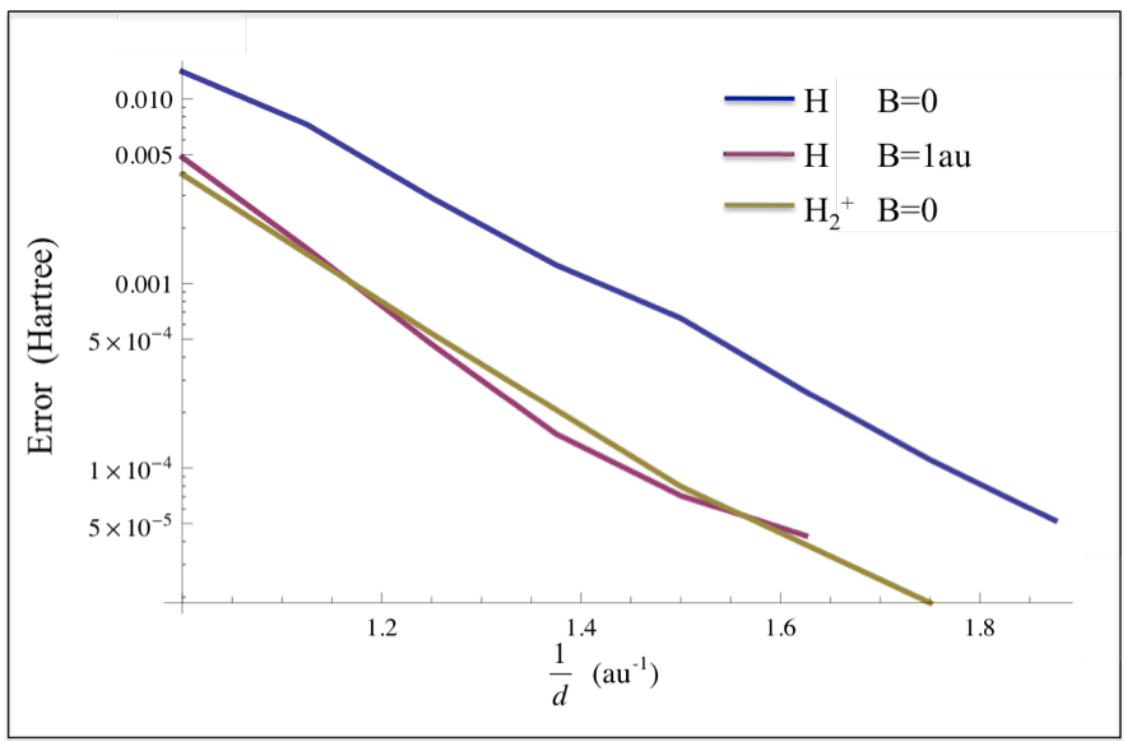

Fig. 2. The convergence of the ground state energy of systems with and without magnetic fields. a) Hydrogen at zero field; the answer is converging towards the analytically known energy of -0.500000. b) Hydrogen in 1 au magnetic field: the true energy is not known well enough to establish a linear convergence plot, but there is rough convergence to -0.3312 Hartree, Thirumalai et. al. [33]. c) we also include $H_2^+$ in zero-field. The convergence approaches the true energy of -0.602634, Scott et. al. [34].

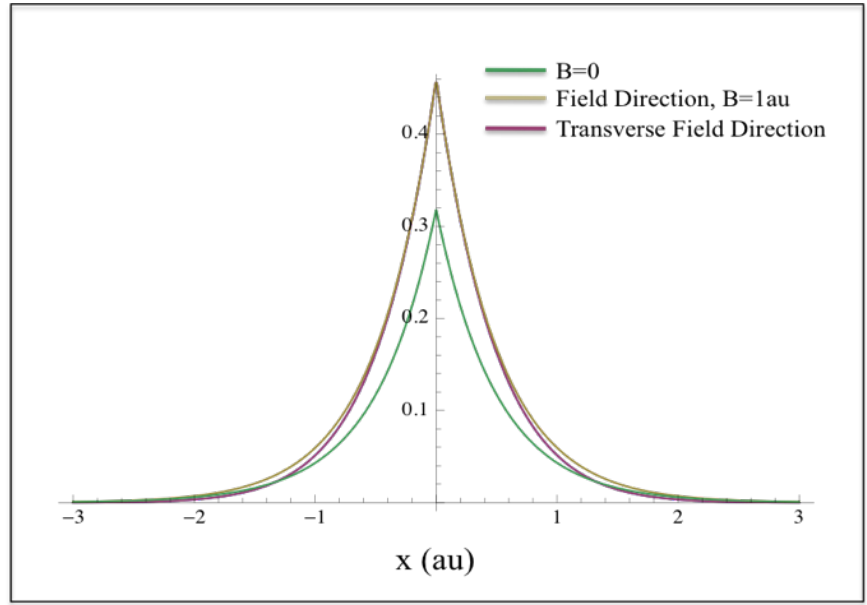

Fig. 3. The cross sections of the ground state Hydrogen wave functions in a B=1au magnetic field using a Gaussian-*Sinc* basis. The zero magnetic field solution is drawn in green. The x-y verses z components are slightly compressed. This calculation was generated with d =0.533 au and a box size of 7.5 au. For reference 1 au = 2.35052 x $10^5$ Tesla.

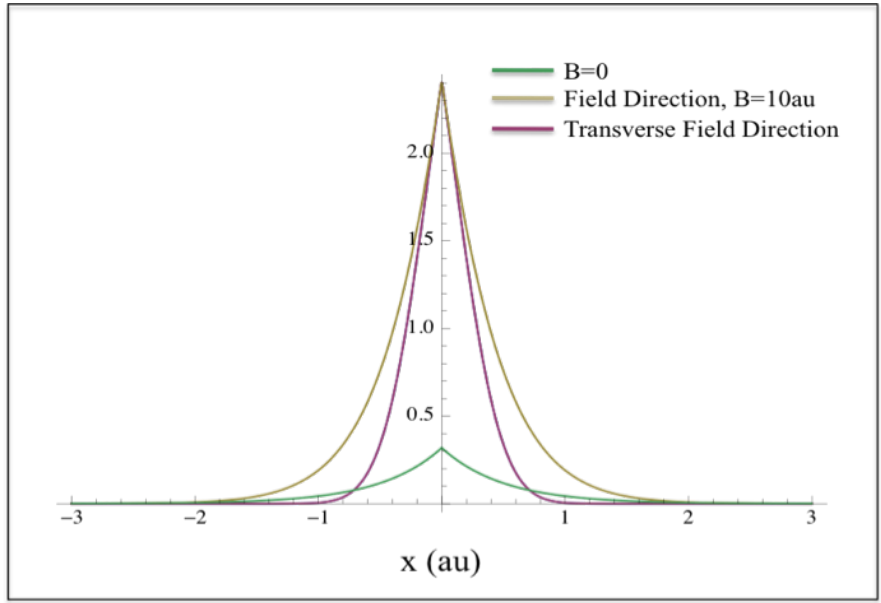

Fig. 4. The cross sections of the ground state Hydrogen wave function in a B=10 au magnetic field using a Gaussian-*Sinc* basis. The zero magnetic field solution is drawn in green. The x-y verses z components are highly compressed. This figure was generated with d=0.300 and a box size of 4.21 au. For reference 1 au = 2.35052 x $10^5$ Tesla

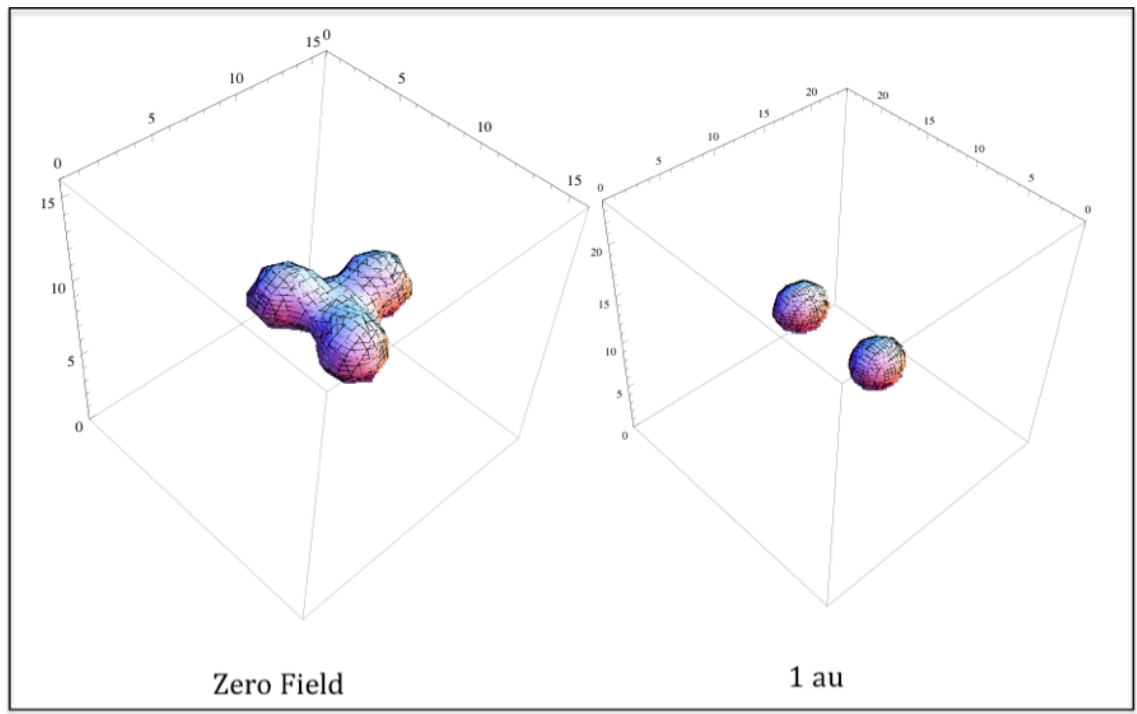

Fig. 5. a) Trigonal electron wave function of $H_3^{2+}$ using a Gaussian-*Sinc* Basis. The three protons are distributed in an equilateral triangle with length 5.25 au. The energy is -0.34 au. The lattice spacing is 1 au. b) Trigonal electron wave function of $H_3^{2+}$ in B=1 au magnetic field. This is the same proton configuration as is shown in figure 6a. The lattice spacing is halved, but box size is unchanged. The third proton receives almost no electron density. The size of the electron wave function decreases with increasing magnetic field. For reference 1 au = 2.35052 x $10^5$ Tesla

Fig. 6. The energies of $H_3^{2+}$ and $H_2^+$ with respect to the energy of Hydrogen in the presence of a magnetic field using the Gaussian-Sinc basis set..

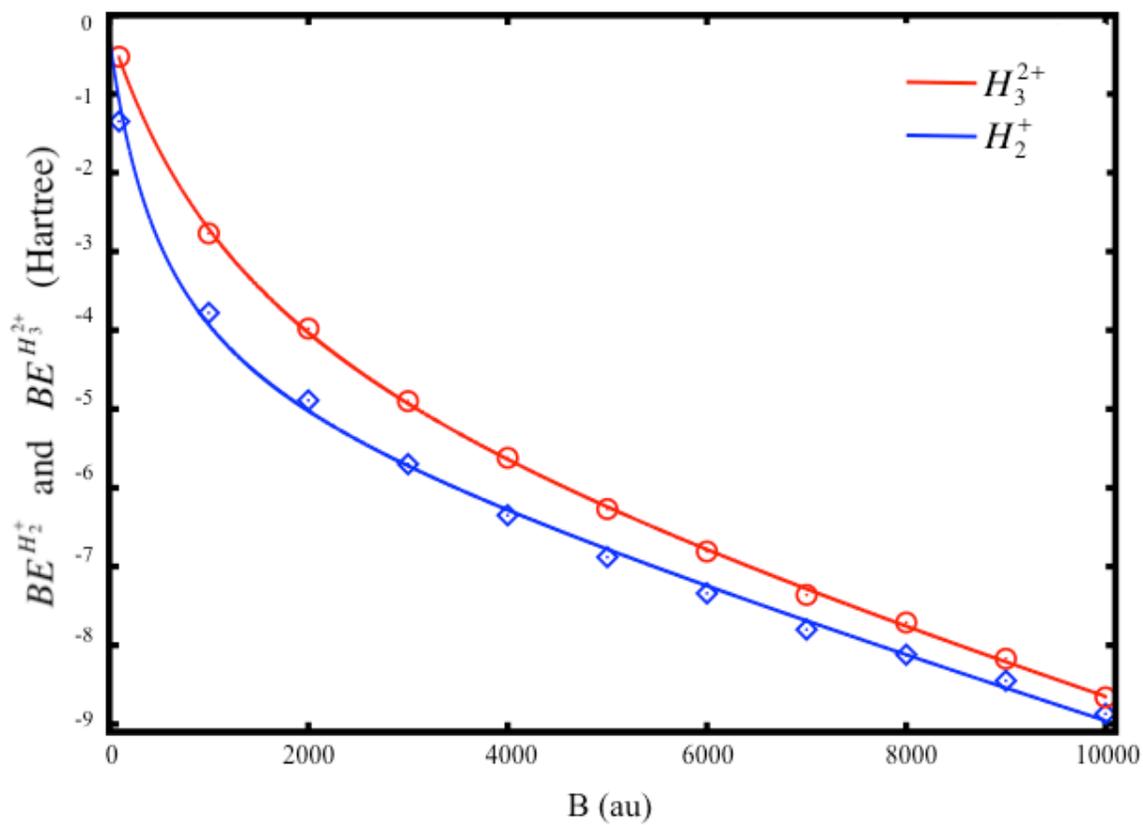

## IX. TABLES

**TABLE I:** Hydrogen in the *Sinc* basis for decreasing grid spacing with a radius of 7.5 au. The number of basis elements on the three grids of decreasing quality are 14,000, 1700, and 250 respectively.

| N | d (au) | T (Hartree) | V (Hartree) | E (Hartree) |
|---|---|---|---|---|
| 250 | 2.000 | 0.284773 | -0.702619 | -0.417846 |
| 1700 | 1.000 | 0.432197 | -0.911736 | -0.479539 |
| 14,000 | 0.500 | 0.487840 | -0.984379 | -0.496604 |

**TABLE II:** Results under translations with increasing *Sinc* basis set range. All calculations are conducted with a grid of d = 1 au. We find the translational quality is dependent only on the edge effects that reduce with an increasing range of the *Sinc* basis, as expected.

| Range (au) | Energy (Hartree) | Energy (½ d shift) |
|---|---|---|
| 3.0 | -0.397029 | -0.416588 |
| 5.0 | -0.475833 | -0.475425 |
| 7.0 | -0.479425 | -0.479045 |
| 9.0 | -0.479587 | -0.479263 |
| 11.0 | -0.479595 | -0.479324 |
| 13.0 | -0.479597 | -0.479364 |

**TABLE III:** The Hydrogen Ground State using a mixed Gaussian-*Sinc* basis. We calculate the core part of the wave function using 3, 6 and 9 s-type Gaussians where we use the equation (5.1) and optimize the variables. The deviation against the correct energies is shown to be exponentially convergent in Figure 3.

| No. of Gaussian | Grid Spacing d (au) | $\alpha_0$ | $\beta_0$ | Condition Number | Energy |
|---|---|---|---|---|---|
| 3 | 2.0 | 0.53 | 4.8 | 105 | -0.498528 |
| 6 | 2.0 | 0.30 | 3.2 | 1905 | -0.499967 |
| 9 | 2.0 | 0.25 | 2.7 | 8700 | -0.499998 |
| 3 | 1.0 | 1.90 | 4.8 | 146 | -0.499831 |
| 6 | 1.0 | 1.15 | 3.2 | 2386 | -0.499995 |
| 9 | 1.0 | 1.05 | 2.7 | 6995 | -0.499999 |
| 3 | 0.5 | 7.1 | 4.8 | 180 | -0.499978 |
| 6 | 0.5 | 5.6 | 3.3 | 960 | -0.499997 |
| 9 | 0.5 | 5.0 | 3.2 | 910 | -0.499999 |

**TABLE IV:** The convergence of Hydrogen in a magnetic field of B=1 au. For reference 1 au = $2.35052 \times 10^5$ Tesla

| d (au) | Range (au) | E (Hartree) |
|---|---|---|
| 1.00 | 4.00 | -0.32638 |
| 0.89 | 4.50 | -0.32965 |
| 0.80 | 5.00 | -0.33073 |
| 0.73 | 5.50 | -0.33105 |
| 0.67 | 6.00 | -0.33113 |
| 0.62 | 6.50 | -0.33116 |
| 0.57 | 7.00 | -0.33116 |
| 0.53 | 7.50 | -0.33116 |

**TABLE V:** Convergence of Hydrogen in a B=10 au magnetic field. For reference 1 au = $2.35052 \times 10^5$ Tesla.

| d (au) | Range (au) | E(Hartree) |
|---|---|---|
| 0.56 | 2.25 | 3.18700 |
| 0.50 | 2.53 | 3.24029 |
| 0.45 | 2.81 | 3.24826 |
| 0.41 | 3.09 | 3.25164 |
| 0.37 | 3.37 | 3.25226 |
| 0.35 | 3.66 | 3.25224 |
| 0.32 | 3.94 | 3.25223 |
| 0.30 | 4.22 | 3.25222 |

**TABLE VI:** Convergence of $H_2^+$ at equilibrium in zero magnetic field.

| d (au) | Range (au) | E (Hartree) |
|---|---|---|
| 1.00 | 4.00 | -0.598735 |
| 0.80 | 5.00 | -0.602098 |
| 0.67 | 6.00 | -0.602555 |
| 0.57 | 7.00 | -0.602616 |

**TABLE VII:** The calculation of $H_2^+$ without a magnetic field under all angles and translations. The calculations are completed with the grid centered on the atom and the grid translated by 0.5 au. Each line was independently scanned and shown to be identical within tolerance.

| B= 0 au | | | | | | | | |
|---|---|---|---|---|---|---|---|---|
| Not translated | | | Translated | | | Reference | | |
| θ | $E_T$ | $R_{eq}$ | θ | $E_T$ | $R_{eq}$ | θ | $E_T$ | $R_{eq}$ |
| 0° | -0.6026 | 2.00 | 0° | -0.6026 | 2.00 | 0° | -0.60263 | 2.00 |
| 45° | -0.6026 | 2.00 | 45° | -0.6026 | 2.00 | 45° | -0.60263 | 2.00 |
| 90° | -0.6026 | 2.00 | 90° | -0.6026 | 2.00 | 90° | -0.60263 | 2.00 |

**TABLE VIII:** A calculation of a translated and origin centered $H_2^+$ in the presence of magnetic fields. The total binding energy, $E_T$, equilibrium distance $R_{eq}$ at the specified angle θ. Comparison is made with reference [16]. For reference 1 au = 2.35052 x $10^5$ Tesla

| B= 1.0 au | | | | | | | | |
|---|---|---|---|---|---|---|---|---|
| Not translated | | | Translated | | | Reference | | |
| θ | $E_T$ | $R_{eq}$ | θ | $E_T$ | $R_{eq}$ | θ | $E_T$ | $R_{eq}$ |
| 0° | -0.4750 | 1.75 | 0° | -0.4750 | 1.76 | 0° | -0.47496 | 1.75 |
| 45° | -0.4623 | 1.69 | 45° | -0.4622 | 1.69 | 45° | -0.45925 | 1.67 |
| 90° | -0.4506 | 1.64 | 90° | -0.4506 | 1.64 | 90° | -0.44956 | 1.63 |
| B=10.0 au | | | | | | | | |
| Not translated | | | Translated | | | Reference | | |
| θ | $E_T$ | $R_{eq}$ | θ | $E_T$ | $R_{eq}$ | θ | $E_T$ | $R_{eq}$ |
| 0° | 2.8251 | 0.96 | 0° | 2.8289 | 0.97 | 0° | 2.82512 | 0.96 |
| 45° | 2.9900 | 0.84 | 45° | 2.9927 | 0.84 | 45° | 3.01165 | 0.81 |
| 90° | 3.1126 | 0.80 | 90° | 3.1160 | 0.80 | 90° | 3.11585 | 0.77 |

**TABLE IX:** Results for Hydrogen, the $H_2^+$ ion and the $H_3^{2+}$ ion in magnetic fields up to 10,000 au. Results are compared to known calculation from references [16].

| B (au) | Energy (Hartree) | | | Reference Energy (Hartree) | | |
|---|---|---|---|---|---|---|
| | H | $H_2^+$ | $H_3^{2+}$ | H | $H_2^+$ | $H_3^{2+}$ |
| 1 | -0.3311 | -0.4750 | - | -0.3312 | -0.47496 | - |
| 10 | 3.2522 | 2.8251 | - | 3.25225 | 2.82512 | 3.3042 |
| 100 | 46.21 | 44.86 | 45.68 | 46.2104 | 44.8545 | 45.6806 |
| 1000 | 492.34 | 488.56 | 489.57 | 492.341 | 488.611 | 489.61 |
| 2000 | 990.69 | 985.80 | 986.71 | | | |
| 3000 | 1489.62 | 1483.92 | 1484.72 | | | |
| 4000 | 1988.81 | 1982.46 | 1983.19 | | | |
| 5000 | 2488.15 | 2481.27 | 2481.88 | | | |
| 6000 | 2987.59 | 2980.25 | 2980.78 | | | |
| 7000 | 3487.16 | 3479.36 | 3479.80 | | | |
| 8000 | 3986.62 | 3978.50 | 3978.91 | | | |
| 9000 | 4486.22 | 4477.77 | 4478.05 | | | |
| 10000 | 4985.97 | 4977.10 | 4977.31 | 4986.02 | 4977.10 | 4977.30 |

# X. REFFERENCES